\documentclass[12pt]{article}
\usepackage{epsfig}
\usepackage{graphicx}
\usepackage{amsmath}
\usepackage{hhline}
\usepackage{amssymb}
\usepackage{times}
\usepackage{cite}

\newlength{\dinwidth}
\newlength{\dinmargin}
\begin{document}  
 


\begin{flushleft}
{\tt \today } \\
\end{flushleft}
\begin{center}
\begin{Large}
{\boldmath \bf Cosmic rays and changes in atmospheric infra-red transmission} \\

\end{Large}
 
\begin{flushleft}
A.D. Erlykin (Lebedev Physical Institute, Moscow and Physics Department, Durham University)\\
T. Sloan 
(Physics Department, Lancaster University)\\
A.W. Wolfendale 
(Physics Department, Durham University)\\

 
\end{flushleft}
\end{center}


\begin{abstract}
\noindent

Recent work by Aplin and Lockwood \cite{AL} was interpreted by them 
as showing that there is a multiplying ratio of order 
10$^{12}$ for the infra-red energy absorbed in the ionization produced 
by cosmic rays in the  atmosphere to the energy content of the cosmic rays 
themselves.  
We argue here that the interpretation of the result in
terms of infra-red absorption by ionization is incorrect and that the
result is therefore most likely due to a technical artefact
 
\end{abstract}

\section{Introduction}

   Atmospheric molecular cluster ions (MCI) are bipolar charged species  
formed by ionization in the atmosphere. The absorption of infra-red 
radiation (IR) by such clusters is interesting since it could have an 
effect on the Earth's radiation budget and thereby allow the ionization 
from cosmic rays (CR) to affect the climate.  Recently, an experiment has 
been described by Aplin and Lockwood (AL) in which they claim to observe   
a large absorption of IR by MCI produced by CR in the atmosphere \cite{AL}. 

In the AL experiment infra-red (IR) detectors are operated close to a small 
CR telescope. The IR band studied is 9.15$\pm$0.45 $\mu$m, a region of 
reduced absorption by atmospheric greenhouse gases \cite{PandO}.  
They observe an average 
decrease of $\sim$2.5 mW/m$^2$ in intensity over this 
wavelength range in a time duration of order 800 seconds 
following counts in the telescope. They assume that the decrease is 
caused by the absorption of IR radiation by MCI  
produced by CR showers, one particle of which gives the 
detected count (usually a muon). They claim that 
the ratio of the total IR energy absorbed by these showers to the 
energy in the CR itself is of order of 10$^{12}$.

This quite remarkable result needs careful independent analysis and this 
is what we propose to do. We will show that the interpretation of 
result as absorption of IR by MCI leads to impossible 
consequences and we conclude that this interpretation is wrong.   

\section{The reasons for believing that the AL interpretation is wrong} 

\subsection{Most AL triggers are from low multiplicity events}

AL propose that the absorption which they observe is from CR 
showers in the upper atmosphere. Their trigger 
is unselective and so they sample all primary CR energies. The 
energy spectrum of CR primaries falls roughly as $E^{-3}$ so their 
triggers (mostly muons) will come mainly from low energy primaries. 
A calculation shows that the average primary energy sampled by their 
trigger is $\sim$ 12 GeV interacting at an altitude between 10 and 20 km 
\cite{Ilya}. The average multiplicity 
of secondary particles at this primary energy will be of 
order 10 \cite{PDG}. The mean transverse momenta of the secondary tracks 
will be of order 0.5 GeV/c \cite{PDG}. Together with the effects of 
multiple Coulomb scattering, this will spread the secondary particles 
over a radius of several hundred metres at the Earth's surface. 
There will be considerable fluctuations about these values but these 
will serve for the order of magnitude estimates we make here.     

From this one sees that the majority of the CR triggers in the AL 
apparatus come from small low energy showers rather than the large 
high energy showers which they assume. A single low energy shower
produces only a small instantaneous increase of the ion pair concentration
in the atmospheric column above their IR detectors, as described below.

\subsection{The observed absorption is inconsistent with laboratory 
measurements}

\label{w1}

This is illustrated by a simple order of magnitude calculation.  
AL draw attention to the measurements of \cite{Carlon, AandM}. These show 
that laboratory measurements  
give rise to absorptions of 1-3\% in two bands centred on wavelengths 
9.15 and 12.3 $\mu$m with MCI columnar concentrations of 10$^{13}$m$^{-2}$. 

CR muons deposit energy at the rate of 1.8 MeV per g cm$^{-2}$ in the 
air \cite{PDG}. The energy 
expended to create an ion pair is 35 eV \cite{PDG}. So each muon 
produces 5.1~10$^4$ ion pairs per gm cm$^{-2}$ (i.e. 66 ion pairs per cm 
of air at ground level). A muon passing through 
the troposphere (lower 700 g/cm$^2$ of the atmosphere) will therefore 
produce 3.6~10$^7$ ion pairs. Let us assume that each muon is, on average,
accompanied 
by of order 10 further muons over an area of order 100 m$^2$. This implies 
an ion columnar density of order 3.6~10$^6$ ion pairs per m$^2$.    

Fig 2 in the AL paper shows that the mean daily 
IR intensity is 350 W/m$^2$ in their broad band detector. The 
intensity in the region of their narrow band detector 
($9.15\pm0.45 \mu$m) will be approximately 5.5\% of this figure 
i.e. 19 W/m$^2$. We make the conservative assumption that all 
this energy flux comes from the top of the troposphere. 

From the laboratory measurements one would deduce, assuming 
that each ion of the pair produces a MCI (i.e. 2 MCI per ion pair), 
that the ionization from CR should absorb 
$2\cdot (0.01-0.03) \cdot 19 \cdot3.6~10^6/10^{13}$ 
i.e. 0.14-0.41 $\mu$W/m$^2$. 
This is 4 orders of magnitude smaller than AL actually observe. 
This absorbed energy is an overestimate since it 
assumes every IR photon passes through the column of ions in the 
shower. In fact only a fraction $\Delta\Omega/4\pi$ of the photons 
will pass through the column where $\Delta \Omega$ is the solid angle 
subtended by the shower at the IR detector. Hence the 
absorption should be even smaller than this estimate. 
  
Furthermore, the time variation of the amplitude decrease seen by AL 
is incompatible with absorption by MCI. The MCI concentration should  
decay exponentially after formation with a time constant of order 
of their lifetime due to recombination. This lifetime could 
be as short as 50 seconds \cite{Mason} but a more modern clculation  
would increase this to of order 500 seconds. In contrast, AL observe 
that the amplitude of their signal actually increases rather than 
decreases with time for 500-700 seconds and then decreases rapidly. 
Hence the time variation observed by AL is not an exponential decay 
and is therefore incompatible with the absorption of IR by MCIs.    
    
In conclusion the magnitude of the AL signal
is inconsistent with their laboratory measurements, and the time
characteristics of the AL signal are inconsistent with those expected 
from the aborption of IR by ions produced by a CR shower. 


\subsection{Implied energy imbalance}

The AL multiplying factor of 10$^{12}$ should be seen against the 
fact that the total sunlight energy density is about 10$^8$ times 
that in CR (adopting the usual CR energy density of 0.5 eV cm$^{-3}$ 
\cite{PDG}).   
Their trigger is unselective and is sensitive to all muons which pass 
through its active solid angle. Hence, on average, each muon must 
behave in a similar way and the effect they observe must therefore 
be cumulative and linear. The implication is that their factor of 
10$^{12}$ then applies, on average, to all CR hitting the Earth.  
Hence the claimed absorption of IR energy by MCI from CR is of 
order 10$^3$ times the total from sunlight falling on Earth (assuming 
on average 10 muons per shower).  

Hence, as well as the inconsistencies described in section \ref{w1}, 
the attenuation which AL claim to measure is also inconsistent with 
conservation of energy. Therefore, their interpretation of the result 
as attenuation of IR by ionization from CR must be wrong. 


\section{Consequences of the result being true}

\subsection{The absorption cross section for IR photons by multi cluster ions}

\subsubsection{The signal from a single muon}
\label{single}

The laboratory measurements of  \cite{Carlon, AandM} imply a measured cross 
section per MCI for absorption of IR photons of 1-3~10$^{-11}$ cm$^2$.   
A comparison is now made with the cross sections which can be deduced 
from the measured attenuations by AL assuming that it comes from absorption 
of IR by MCI produced by ionization from CR particles.    

The probability of an IR photon to be directed towards the AL detector 
and to be absorbed by MCIs from a single ionizing track is given by 
geometry to be  
\begin{equation} 
P= \frac{I \sigma}{4 \pi a}(\alpha_1 - \alpha_2 +\frac{1}{2}\sin2\alpha_1-\frac{1}{2}\sin2\alpha_2). 
\label{one}
\end{equation}
This equation is derived in the Appendix. 
Here $I$ is the number of MCI per unit length of the track, $a$ is the 
perpendicular distance from a projection of the track to the IR detector 
and $\sigma$ is the absorption cross section for an IR photon by a MCI. 
The angles $\alpha_1$ and $\alpha_2$ (see figure \ref{fig1}) are those 
between the line in the plane 
of the track through the detector perpendicular to the track projection 
and the line in the same plane from the detector to the start and end 
points of the track, respectively. 

It can be seen from equation \ref{one} that the absorption probability 
decreases linearly with the perpendicular distance, $a$, of the 
projection of the particle track to the detector. Hence the closest 
tracks to the detector are the most important ones for IR absorption.  
It can also be seen that for tracks which begin and end at high 
altitude the difference between the angles $\alpha_1$ and  $\alpha_2$ 
will be small and therefore 
the absorption probability for such tracks is small. Hence, the 
contribution from high altitude absorption will be small except for the 
rather rare extensive air showers from very high energy primaries 
which produce large numbers of particles.  Such events are rare since 
the primary CR spectrum falls roughly as $E^{-2.6}$ \cite{Wolfendale}, 
where $E$ is the primary energy.  They are considered separately in 
section \ref{showers}. For a single muon, the quantity $I$ will fall 
as the altitude increases due to the reduction of pressure with altitude. 
This is partly offset, however, by the increased ionization from the 
few other secondary tracks associated with the detected muon \cite{Ilya}. 
In fact, the decreasing rate of change of the angle 
$\alpha$ with altitude implies that most of the absorption takes place 
in the vicinity of the detector, so that the changes in $I$ will be 
insignificant.   

The absorption probability measured from the AL experiment is difficult 
to estimate precisely. However, rough order of magnitude estimates are 
possible as follows.   
Assuming that the principal source of IR is 
radiation from the lower atmosphere, the total source energy in their 
wavelength range will be 19 W/m$^2$ (350 W total with a fraction 0.055 
in their wavelength range). If, however, the 
source is mainly radiation from the stratosphere, the total will be 
lower, implying a higher absorption probability (higher cross section) . 
To obtain a conservative lower limit on the cross section we take the 
measured probability to be the ratio of the observed absorption of 
2.5 mW/m$^2$ to the estimated source energy of 19 W/m$^2$ i.e. 
1.3~10$^{-4}$, the smaller of the two probabilities.         


The columnar density of MCI is computed from the rate of production 
of ionization by muons (see above) assuming that 
each ion pair produces 
a MCI. The absorption cross section is then computed from the AL observed 
attenuation and the density of MCI production as follows.  

From equation \ref{one} 
the absorption will be dominated by the track closest to the detector 
which in the majority of cases will be the trigger muon.  
In this case the angle $\alpha_1$ is almost 
$\pi/2$ radians and the angle $\alpha_2$ will be the angle of the 
muon track to the vertical which is usually small since the muon 
angular distribution peaks around the vertical direction \cite{Wolfendale}.  
Substituting these values into equation \ref{one},   
the absorption cross section for IR photons 
will then be of order 3~10$^{-3}$ cm$^2$ per MCI. Allowing 
for IR photons produced below the level of the track or the fact 
that not every ion pair produces a MCI
 will increase the value of this estimated cross section. Hence this 
value is a lower limit on the cross section necessary to satisfy 
the AL observations.  
This lower limit is 8 orders of magnitude greater than the measured 
cross section \cite{Carlon, AandM}.

We have attempted to be conservative to find this lower
limit on the cross section for absorption of an IR photon 
by MCI.  Some of the numbers are debatable 
and adjustments to the numbers could be made which may    
decrease it by a small factor. However, it will be impossible 
to reduce the implied cross section by the 8 orders of magnitude 
needed to be compatible with the laboratory measurements. 

Hence, the cross section implied by the AL measurements is again 
incompatible with the laboratory measurements and it is also 
unphysically large for a molecular process.   

\subsubsection{The signal from CR showers}
\label{showers}

AL \cite{AL2} propose that the majority of the absorption is by showers 
at high altitude. Such showers dissipate most of their energy 
as lower energy secondary particles at altitudes 
between 5 to 15 km. Hence the term in 
$\alpha_1-\alpha_2$ is of order 10$^{-3}$ for such particles, 
assuming that the value of $a\sim 300$ cm. The atmospheric 
pressure at this altitude is roughly 1/3 that at ground level. Hence the 
number of ion pairs will be of order 20 per cm per secondary particle. 
We assume that the 
cross section for the absorption of IR photons is the measured value 
of  2 10$^{-11}$ cm$^2$ \cite{Carlon, AandM}.
Substituting these values into equation \ref{one} shows that the probability 
for the absorption of IR photons at this altitude is of order 10$^{-16}$ 
per secondary 
particle. We showed above that the total absorption probability 
for IR photons implied by the AL measurements is of order 10$^{-4}$. 
Hence one needs showers containing of order 10$^{12}$ particles to 
produce the absorption which they observe. 

 
Extrapolating from shower measurements at lower energies, the 
primary particle energy needed to produce this number of secondary 
particles is of order 10$^{12}$ GeV. 
Such an energy CR primary is greater than the maximum energy currently 
being observed and these events are very rare indeed, with fluxes of 
order 1 per square kilometre per century\cite{Auger}.  This is 
too rare to influence the average from randomly selected muons. 

AL imply that their result could be due to absorption in CR showers 
at high altitude \cite{AL2}.  As we show above in section \ref{single} 
most of the absorption should, instead, be attributable to single or 
small numbers of particles passing close to the IR detector.  
Events which give large enough numbers of particles to 
produce significant absorption in the upper atmosphere are 
extremely rare as shown above. Thus large CR showers cannot be 
responsible for the AL result.  


\subsection{Implications of the detected signals}

Yet another way of looking at the consequence of the signal in 
the AL experiment being true is simply to consider the contribution 
to the IR absorption of all the other CR muons which arrive within 
the temporal and spatial window of the IR detector. 

Equation \ref{one} shows that most of the absorption occurs from muon tracks 
within a few metres of the IR detector. The total 
number of muons in 1 second passing through a disk of radius $R$ 
is $\sim\pi R^2 I_{CR}$ where 
$I_{CR}\sim$80 m$^{-2}$sr$^{-1}$s$^{-1}$\cite{Wolfendale}
is the observed vertical CR muon rate at the Earth's surface. This 
gives a total of 6300 muons passing through a disk of radius 5m each 
second.  According to the AL measurements each 
muon absorbs $\sim$2.5 mJ in each second for a time of 800 seconds. So 
the total IR energy absorbed in any 1 second by muons in the vicinity 
of their detector is $\sim0.0025\cdot6300\cdot800~\sim$ 12600 J i.e. 
the total IR power which would be absorbed is of order  
12.6 kW in the 9$\mu$m band. However, there are only $\sim$350 W of power 
available over the whole IR spectrum. Hence again the result of the AL 
experiment implies an unphysical value.  

\section{Likely explanations and conclusions}

We have demonstrated that the results of the AL measurements, as 
interpreted by AL, lead to impossible consequences. 
What then could be the reason for the result? 

It might be thought that an explanation is that it is due to 
some new unknown process. This seems highly unlikely since the 
contributing processes involve rather low energy electromagnetism. 
Furthermore, there would still be the inconsistency with their own 
laboratory measurements. 
 

 
A more likely explanation is that the result is due to a bias  
or 'cross-talk' between the CR and IR detectors. Averaging noisy signals 
to produce a small observed deviation from zero such as is done in 
the AL experiment is very sensitive either to the presence of an 
apparatus bias or to such cross talk. 

It is evident that an independent analysis of IR signals associated 
with CR is needed before the dramatic results of AL are considered 
further. Such analysis should include the careful monitoring of 
atmospheric conditions and searches for apparatus biases eg by an equal  
study of random triggers and CR triggers.   

\section{Appendix}

{\bf Probability of absorption of an IR photon by a single ionizing track}

\begin{figure}[ht]
\hspace{-7pc}
\includegraphics[width=28pc]{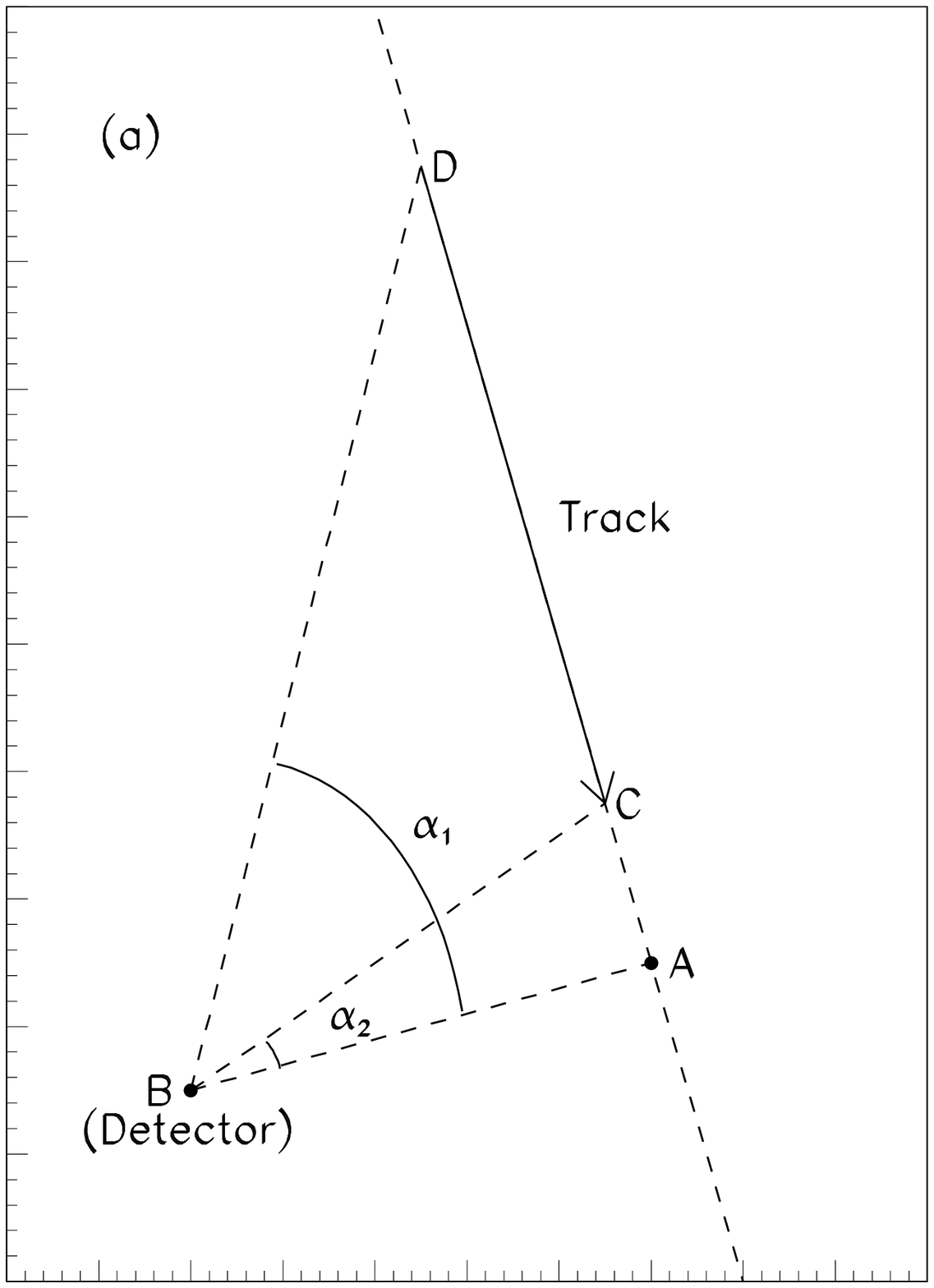}
\hspace{-7pc}
\includegraphics[width=28pc]{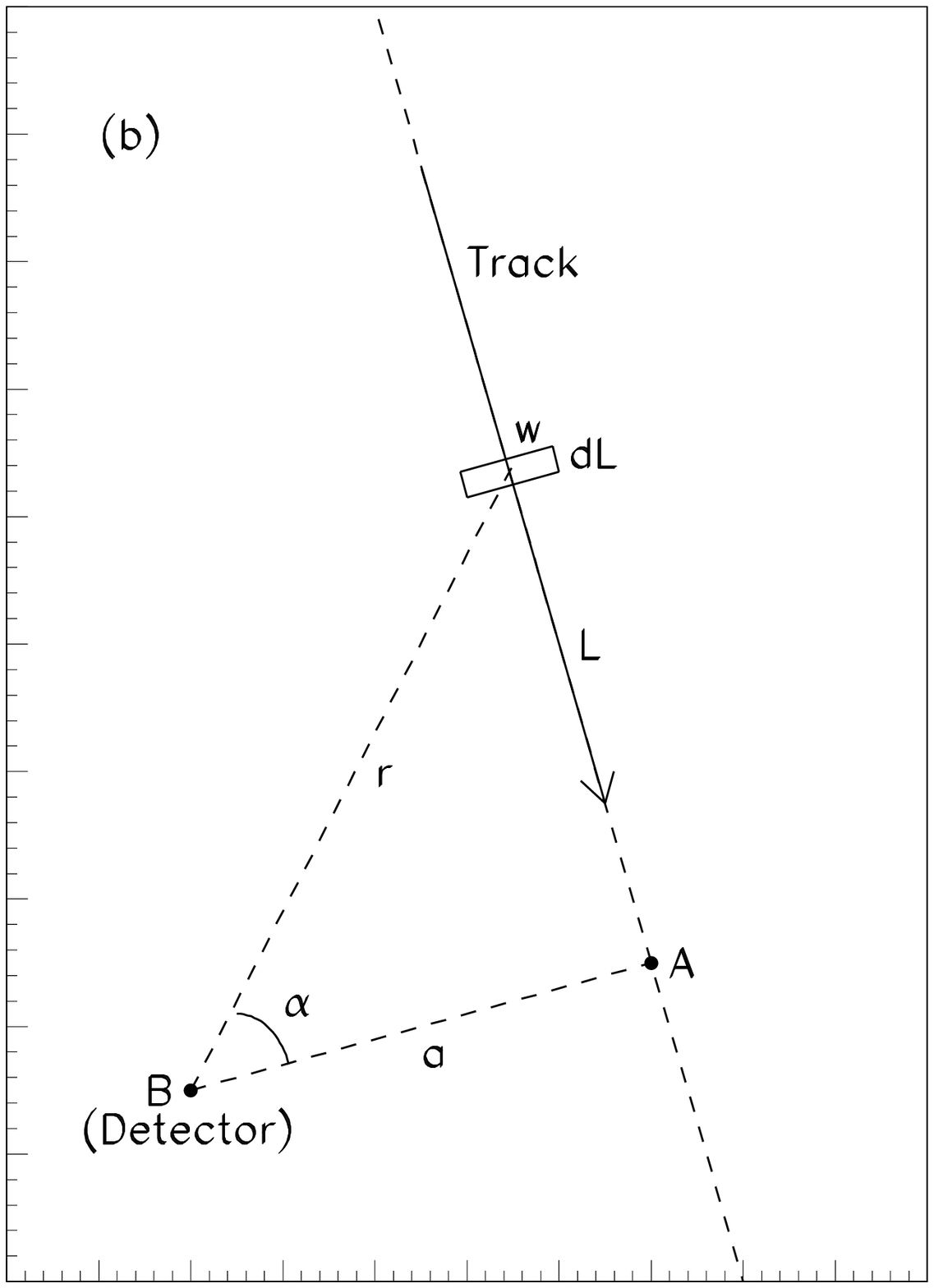}
\caption{\label{fig1} (a) CR track starting at point D and ending 
at point C. The detector is at point B. The line AB is a perpendicular 
from the detector to the line of the track. Angle ABD is $\alpha_1$ 
and ABC is $\alpha_2$. (b) As (a) but showing a small element of the track, 
thickness $dL$, radius $w$ containing the ions. The element is at 
distance $L$ from point A and subtends angle $\alpha$ to the perpendicular 
AB.  } 
\end{figure}

The geometry of the single ionizing track is shown in figure \ref{fig1}.
Take a small element of the track of length $dL$ at distance $L$ from 
the point on the perpendicular between the detector and the track 
(line AB in 
figure \ref{fig1}). This line subtends an angle $\alpha$ to the element.  
Assuming an isotropic distribution of IR photons, the probability of 
a downward-going photon moving in the direction of 
the detector traversing this element of track is $d\Omega/2\pi$ 
where $d\Omega=\pi w^2 \sin \alpha/r^2$ is the solid angle subtended by 
the element at the detector. Here $w$ is the radius 
of the element and $r$ is its distance to the detector. 
The probability that this photon is absorbed in the 
element $dL$ is $dx/\lambda$ where $dx=dL/\sin \alpha$ is the thickness 
of the element traversed and $\lambda=1/\nu \sigma$ is the mean free 
path of the photon in the sea of MCI around the track. The mean density 
of MCI in the element is $\nu=I/\pi w^2$ with $I$ the number of MCI 
per unit length produced by the track and $\sigma$ 
is the absorption cross section for an IR photon by a MCI. 
Here it is assumed that the ions drift 
outwards by Brownian motion to fill a cylindrical column of radius $w$ 
($w >> dL$) 
at a certain time. Hence the probability that the IR photon is 
absorbed in the element is 
\begin{equation} 
dP=\frac{d\Omega}{2\pi} \frac{dx}{\lambda} = \frac{\pi w^2 \sin{\alpha}}{2\pi r^2}\frac{dL}{\lambda\sin{\alpha}} = \frac{I \sigma dL}{2\pi r^2}
\label{eq1}
\end{equation} 
The unknown track radius $w$ cancels so that the value of $dP$ does 
not depend on time.  

Substituting that $L/r=\sin{\alpha}$ so that $dL=r\cos{\alpha}d\alpha$ 
and $a/r=\cos{\alpha}$ gives 
\begin{equation}
dP=\frac{I \sigma}{2 \pi a} \cos^2{\alpha} d\alpha, 
\label{eq2} 
\end{equation} 
here $a$ is the perpendicular distance of the track to the detector 
(see figure \ref{fig1}). 
Integration of equation \ref{eq2} over 
the length of the track (AD in figure \ref{fig1}) i.e. between 
the angular limits $\alpha_1$  and $\alpha_2$ gives the total 
probability for an IR photon to be absorbed by the track which is 
given in equation \ref{one}.  

\section{Acknowledgements}
{We thank the Kohn Foundation for financial support. We also thank an 
unknown referee for helpful comments.}

\begin{thebibliography}{99}  

\bibitem{AL} Aplin, K.L. and Lockwood, M., ``Cosmic ray modulation 
of infra-red radiation in the atmosphere''. Env. Res.Lett., 8 (2013) 015026.

\bibitem{PandO} Peixoto, J.P. and Oort, A.H. ``The Physics of Climate'', 
published by American Institute of Physics (1992), ISBN 0-88318-712-4 
(see figure 6.2).  




\bibitem{Ilya} Usoskin I.G. and Kovaltsov, A., ``Cosmic ray ionization in the 
atmosphere. Full modelling and practical applications'', 
J. Geophys. Res. 111, (2006) D21206.

\bibitem{PDG} For reviews of the behaviour of charged particles in matter 
and of cosmic ray properties see reports of the Particle Data Group,  
Beringer, J. et al., Phys. Rev. D86 (2012) 010001, also available at 
http://pdg.lbl.gov/2013/reviews/contents\_sports.html. See also the 
references therein.

\bibitem{Carlon} Carlon, H., `` Infrared absorption and ion content of 
of moist atmospheric air'', Infrared Physics 22 (1982) 43-49. 

\bibitem{AandM} Aplin, K.L. and McPheat, R.A.,'' Absorption of infra-red 
radiation by atmospheric molecular cluster ions'', J. Atmos. Sol.-Terr. Phys. 
67 (2005) 775-783.

\bibitem{Mason} Mason, B.J., ``The Physics of Clouds'', Clarendon Press, 
Oxford (1971).   

\bibitem{Wolfendale} Wolfendale, A.W. ``Cosmic Rays'', Newnes, London (1963).  


\bibitem{AL2} Aplin, K.L. and Lockwood, M., private communication. 

\bibitem{Auger} Pierre Auger Observatory, Abraham, J. et al., Observation  
of the suppression of the flux of cosmic rays above $4~10^{19}$ eV.  
Phys. Rev. Letts. 101 (2008) 061101. 

\end{thebibliography}
\end{document}